\documentclass[preprint,amsmath,amssymb,prl,superscriptaddress]{revtex4-1}

\usepackage{graphicx}

\begin{document}

\title{Screening and atomic-scale engineering of the potential at a topological insulator surface}

\author{P.~L\"{o}ptien}
\affiliation{Institute for Applied Physics, Universit\"{a}t Hamburg, D-20355 Hamburg, Germany}
\author{L.~Zhou}
 \email[Email: ]{lzhou@physnet.uni-hamburg.de}
\affiliation{Institute for Applied Physics, Universit\"{a}t Hamburg, D-20355 Hamburg, Germany}
\author{J.~Wiebe}
\affiliation{Institute for Applied Physics, Universit\"{a}t Hamburg, D-20355 Hamburg, Germany}
\author{A.A.~Khajetoorians}
\affiliation{Institute for Applied Physics, Universit\"{a}t Hamburg, D-20355 Hamburg, Germany}
\author{J.L.~Mi}
\affiliation{Center for Materials Crystallography, Department of Chemistry, Interdisciplinary Nanoscience Center, Aarhus University, 8000 Aarhus C, Denmark}
\author{B.B.~Iversen}
\affiliation{Center for Materials Crystallography, Department of Chemistry, Interdisciplinary Nanoscience Center, Aarhus University, 8000 Aarhus C, Denmark}
\author{Ph.~Hofmann}
\affiliation{Department of Physics and Astronomy, Interdisciplinary Nanoscience Center, Aarhus University, 8000 Aarhus C, Denmark}
\author{R.~Wiesendanger}
\affiliation{Institute for Applied Physics, Universit\"{a}t Hamburg, D-20355 Hamburg, Germany}

\date{\today}

\begin{abstract}
\noindent
The electrostatic behavior of a prototypical three-dimensional topological insulator Bi$_2$Se$_3$(111) is investigated by a scanning tunneling microscopy (STM) study of the distribution of Rb atoms adsorbed on the surface. The positively charged ions are screened by both free electrons residing in the topological surface state as well as band bending induced quantum well states of the conduction band, leading to a surprisingly short screening length.  Combining a theoretical description of the potential energy with STM-based atomic manipulation, we demonstrate the ability to create tailored electronic potential landscapes on topological surfaces with atomic-scale control.
\end{abstract}


\keywords{topological insulators}

\maketitle

Topological insulators (TIs) belong to a unique class of exotic quantum materials hosting Dirac-dispersing charge carriers with helical spin textures at surfaces, which are topologically protected by time-reversal symmetry \cite{Hasan2010, Qi2011}.
These so-called topological surface states (TSS) have been predicted to host a variety of novel phenomena, thus making them promising for future generation spintronics and quantum computing applications.

However, the ability to harness the topological character of these materials requires the ability to gate the TSS to control surface transport, analogous to other two-dimensional (2D) systems \cite{Balog2010}, and for some of the applications to create a gap at the Dirac point. While extensive effort has been spent on understanding whether magnetism induces a gap at the Dirac point~\cite{Biswas2010, Wray2011, Honolka2012, Schlenk2013}, surface transport through the TSS is still an experimental challenge. This is due in large part to material quality, poor electron mobility, and proper dielectrics. Impurities near the surface can locally gate the TSS leading to unwanted charge disorder \cite{Beidenkopf2011, Adam2012, Mann2013}, which limits charge mobility \cite{Kim2012}. More strikingly, previous studies have thoroughly shown that a dilute amount of impurities on the surface can heavily modify the TSS via band bending, which induces additional quantum well states (QWS) \cite{Bianchi2010} at the surface with a Rashba character~\cite{King2011}.  These results clearly question if 3D TIs with sufficient mobility can be fabricated and how the application of metallic electrodes will modify charge transport through the TSS.

We use scanning tunneling microscopy (STM) to study the screening behavior produced by positively charged Rb atoms on the surface of the prototypical TI Bi$_2$Se$_3$. By analyzing the pair correlation functions~\cite{Hansen1976, Fernandez2007, Song2012}, the screened Coulomb potential between surface Rb atoms is extracted, from which the electrostatic properties of the TI are determined. By varying the surface and bulk doping, we show that the charge screening is mainly provided by 2D electrons, residing in both the TSS and the QWS, resulting in a surprisingly small screening length. 
With that knowledge, we demonstrate the ability to engineer the potential landscape, in which the TSS resides, by atomic manipulation of single Rb atoms on the surface.

Both stoichiometric and Ca-doped Bi$_2$Se$_3$ single crystals were grown and characterized as described in \cite{Bianchi2010, Bianchi2012}. Stoichiometric samples are highly $n$-doped resulting from bulk defects, while Ca doping shifts the Dirac point to lie within 50~meV of the Fermi energy~\cite{Bianchi2012}. All experiments were conducted under ultra-high vacuum conditions with a base pressure below $2 \times 10^{-10}$~mbar. The crystals were cleaved \textit{in-situ} at room temperature exposing the (111) surface. Rubidium was subsequently deposited at room temperature from a commercial dispenser. After deposition, the samples were immediately quenched to low temperatures within the microscope, where STM experiments were performed at $T=4.3$~K \cite{Bianchi2012}. STM topographs were recorded in constant-current mode, with a tunneling current $I_{\mathrm T} = 10$~pA and a sample bias voltage $V_{\mathrm B}= 1$ V. The Rb coverage was varied between 0.6 to 7.3\% of a monolayer (ML), where we define a ML relative to the number of Se atoms in the first layer.  With these sample preparation conditions, Rb atoms remain on the surface and do not intercalate as previously demonstrated, where the electronic properties of Rb-covered Bi$_2$Se$_3$ as a function of annealing was characterized on the same Bi$_2$Se$_3$ crystals used here~\cite{Bianchi2012}.

\begin{figure}[htbp]
\includegraphics[width=0.6\columnwidth]{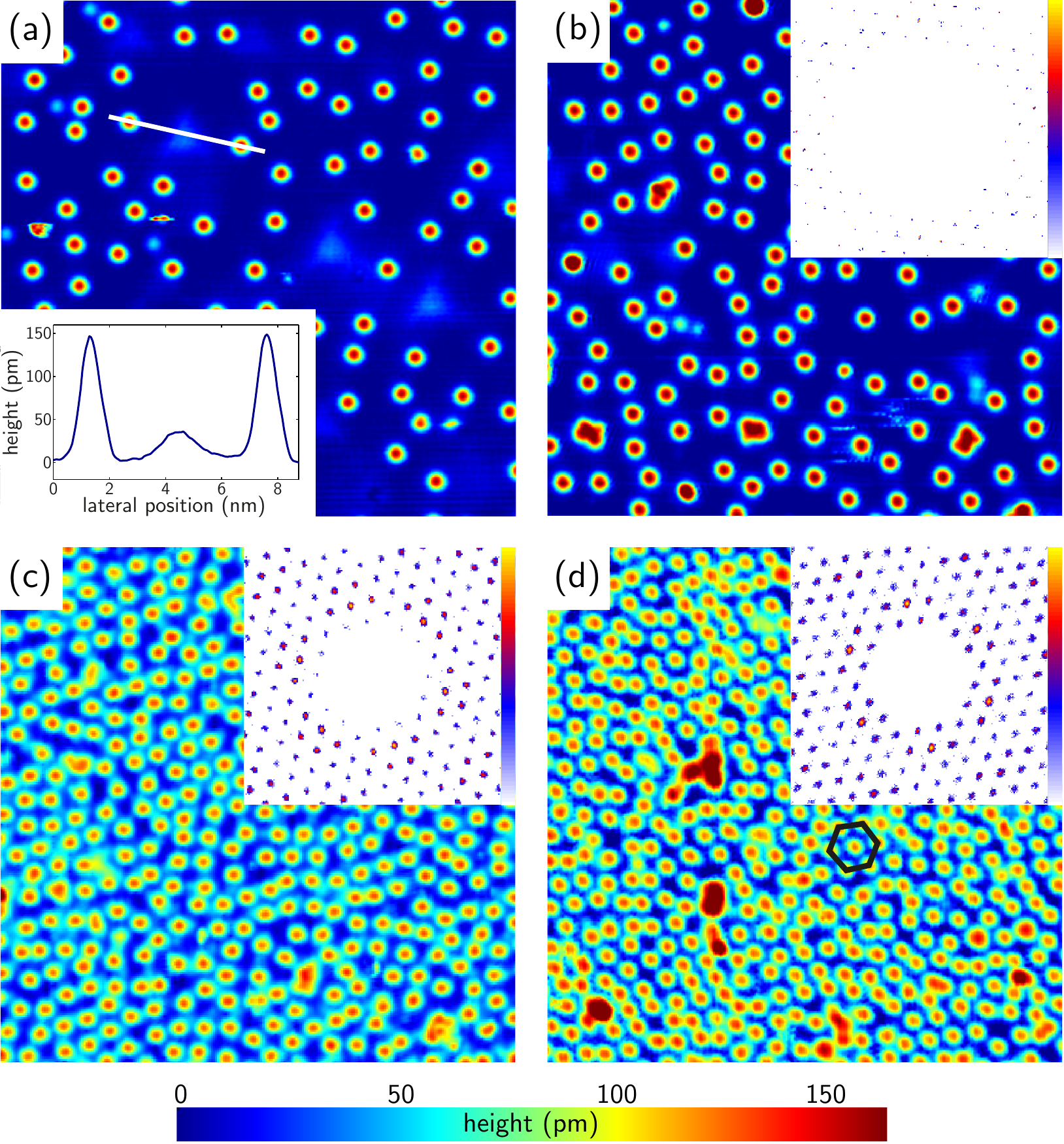}
\caption{\label{Topographies}STM topographs of Rb single atoms on stoichiometric and Ca-doped Bi$_2$Se$_3$ (30 nm $\times$ 30 nm). (a)~Stoichiometric Bi$_2$Se$_3$ with a 1.2\%~ML coverage of Rb. The inset indicates a cross section along the line in (a) that depicts the apparent heights of a Se vacancy and two Rb atoms. (b)~Ca-doped Bi$_2$Se$_3$ with 2.0\% ML Rb. (c) Stoichiometric Bi$_2$Se$_3$ with 5.8\% ML Rb. (d) Stoichiometric Bi$_2$Se$_3$ with 6.9\% ML Rb. The small hexagon depicts the $(\sqrt{12}\times\sqrt{12})\,R30^\circ$ superstructure.
Insets in (b-d): Autocorrelation plots of the atoms' positions (5~nm $\times$ 5~nm; each vector starts at the center. The number of the individual counts is plotted logarithmically using the depicted color scale).}
\end{figure}
STM images of dilutely covered surfaces reveal a distribution of single Rb atoms. These appear as circular protrusions with an apparent height of $\Delta z \approx 150$~pm [Fig.~\ref{Topographies}(a)]. Atomically resolved images of the underlying Bi$_2$Se$_3$(111) substrate show that Rb atoms reside at only one of the two possible surface hollow sites~\cite{Supplement}. In addition, Se subsurface vacancies appear as extended triangular shapes with $\Delta z \approx 30\ \mathrm{pm}$, depending on the depth of the vacancy~\cite{Urazhdin2002a}. A strongly reduced density of Rb atoms in the vicinity of these positively charged Se vacancies, together with the observation that Rb atoms do not cluster, indicates the positive charge of single Rb atoms on the surface \cite{Valla2012,Bianchi2012}. From topographs of the pristine surface with a total area of $\approx$~50,000~nm$^2$ we estimate the surface density of Se vacancies and Ca acceptors to be one to two orders of magnitude smaller than the surface density of Rb atoms. We therefore neglect their effect on the distribution of Rb atoms on the surface.

\begin{figure}[htbp]
	\includegraphics[width=0.6\columnwidth]{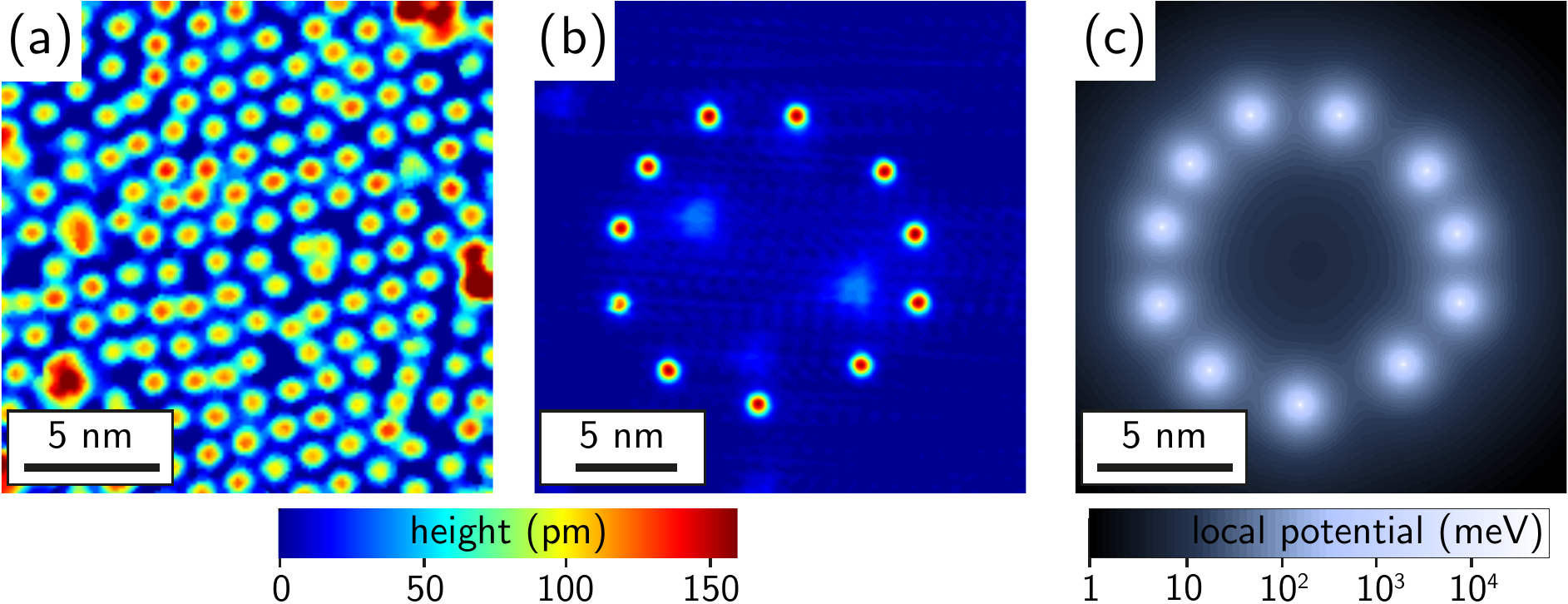}
	\caption{Tailoring the local potential by surface doping with Rb atoms. (a)~Self-assembly of Rb atoms into a superlattice array driven by Coulomb repulsion. (b)~Eleven Rb atoms manipulated with the STM tip. The triangles in the center of the ``corral'' are native subsurface Se vacancies. Manipulation parameters: $I_\mathrm T = 4.6\ \mathrm{nA}$, $V_\mathrm B = 600\ \mathrm{mV}$. (c)~Potential landscapes of (b) generated by using $v_\mathrm{2D}(r)$.}
	\label{Outlook}
\end{figure}

Images with a higher coverage of Rb atoms \mbox{[Fig.~\ref{Topographies}(b-d)]} reveal a very homogeneous, highly non-randomized distribution of Rb atoms across the surface.  In certain regions, ordered superlattice arrays of Rb with a $(\sqrt{12}\times\sqrt{12})\,R30^\circ$ structure can be clearly observed when the Rb coverage is sufficiently high [small hexagon in Fig.~\ref{Topographies}(d) and Fig.~\ref{Outlook}(a)]. Such ordering of impurities on surfaces have been attributed to various substrate-mediated electronic interactions~\cite{Silly2004, Fernandez2007, Song2012}. The positive charge of Rb on the surface suggests this ordering of Rb on the surface is driven by a repulsive Coulombic interaction between Rb atoms. Moreover, the observation of such ordering indicates that upon deposition, the thermal energy is sufficient for the system to equilibrate.

\begin{figure}[t]
	\includegraphics[width=0.6\columnwidth]{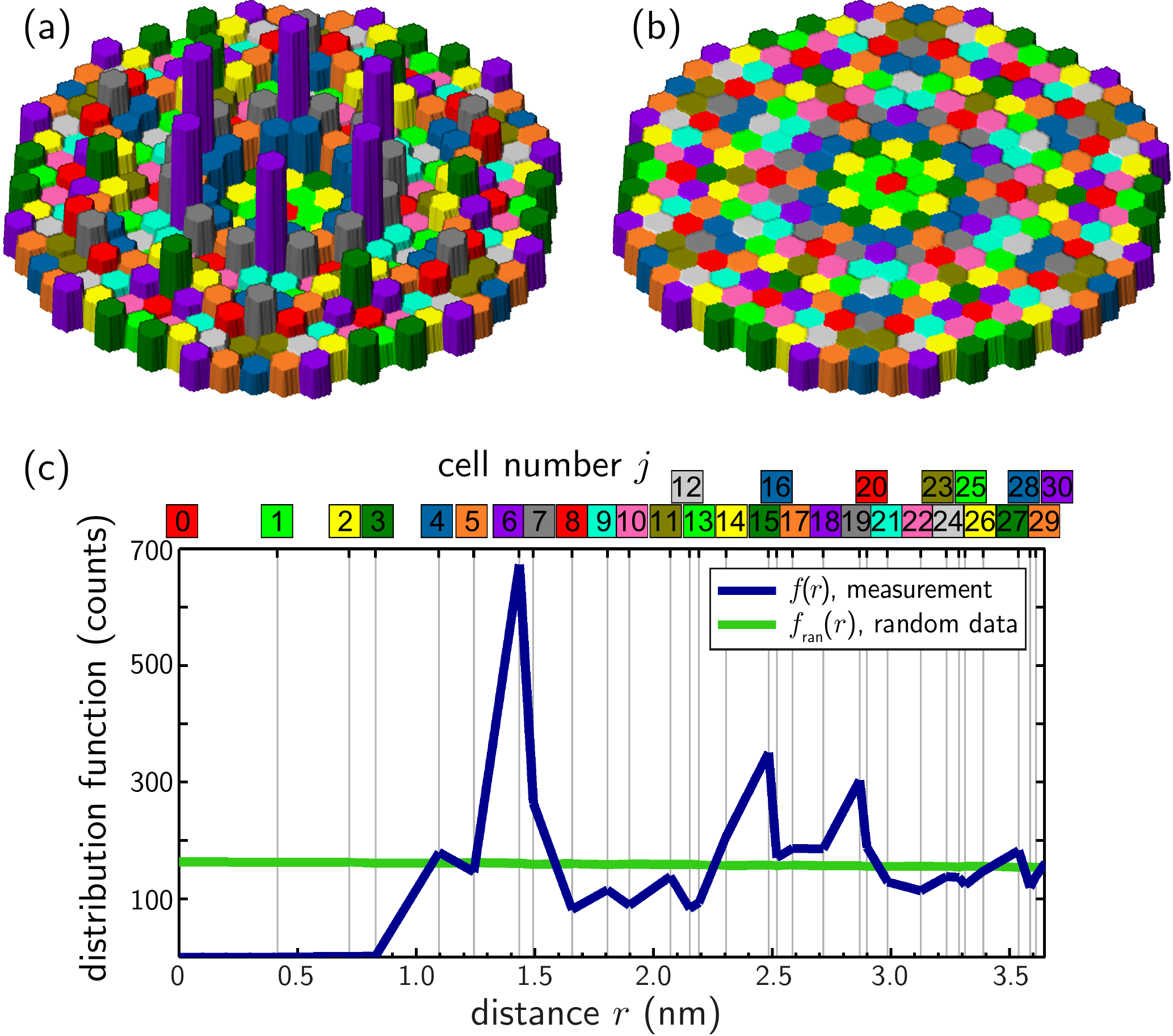}
	\caption{(a) The counts, represented by height, from the autocorrelation images are indexed into colored hexagonal cells labeled by $(j,k)$, shown here for the coverage in Fig.~\ref{Topographies}(d). (b) Same procedure for a randomly generated distribution with the same coverage. (c) Distribution functions $f(r)$ and $f_\mathrm{ran}(r)$ extracted from (a,b), respectively, after averaging over all $k$ values for a given $j$.  Each hexagonal cell $j$ is color-coded in (a,b) as indicated above the graph.}
	\label{f_fran}
\end{figure}
Autocorrelation images \cite{Supplement} of the coverage-dependent distributions of Rb [Fig.~\ref{Topographies}(b-d)], which illustrate the displacement of all surrounding atoms with respect to a given atom, for all atoms in the image, reveal a hexagonal distribution of Rb atoms which reflect the underlying symmetry of the substrate surface lattice. Moreover, these images reveal a radius of $r \approx 1\ \mathrm{nm}$ around every Rb atom at which no other Rb atoms reside.  To quantify this effect and relate it to the underlying potential landscape, we performed a statistical analysis of each Rb atom and its neighbors.  Since the autocorrelation images reveal a non-circular symmetric distribution, we consider a vector-resolved analysis of the displacement of each Rb atom relative to its neighbors~\cite{Trost1996}.  In total, the coordinates of $>$20,000 atoms on an area of $280\ \mathrm{nm}\times 280\ \mathrm{nm}$ were determined from STM topographs.
We define individual bins represented by hexagonal unit cells that resemble the allowed binding sites.
All unit cells at a given distance are indexed by the same parameter $j$, while the parameter $k$ distinguishes the different positions for a given $j$. For a given autocorrelation image, we count the number of atoms in a given unit cell~[Fig.~\ref{f_fran}(a)]. 

There, the highest peaks are observed for cells where \mbox{$j=6$} (purple), which corresponds to the \mbox{$(\sqrt{12}\times\sqrt{12})\,R30^\circ$} superstructure and indicates that Rb atoms have the highest probable interatomic distances of $r = 1.43\ \mathrm{nm}$. 
The subsequent local maxima, namely for $j = 15, 19$ (green, gray), have the same displacement from cells where $j =6$, as cells with $j = 6$ have from the origin. This further indicates the long range nature of the Rb-Rb interaction. Upon closer inspection, the experimental distribution is found to be symmetric, namely independent of the index $k$, thereby allowing all cells with a given $j$ to be averaged over all $k$.  The resultant coverage- and distance-dependent distribution function, $f(r)$, which counts the number of Rb atoms at a given distance from any Rb atom, is graphed in Fig.~\ref{f_fran}(c).  While $f(r)$ reveals information about the potential landscape, it is coverage dependent and needs to be normalized to a random distribution function, $f_\mathrm{ran}(r)$, with the same coverage [Fig.~\ref{f_fran}(b,c)]~\cite{Supplement}. The ratio between both distribution functions yields the pair correlation function $g(r) = f(r)/f_\mathrm{ran}(r)$, which represents the potential of mean force $w(r)$, namely the work required to pull two Rb atoms from infinite distance to a distance $r$~\cite{Hansen1976}:
	\[-\ln g(r) = w(r)/k_\mathrm BT = \left[v(r) + \Delta w(r)\right]/k_\mathrm BT\ .\]
Here, $w(r)$ can be separated into two terms, namely a pairwise term called the pair potential $v(r)$ and a higher order term $\Delta w(r)$.
While $v(r)$ equals $w(r)$ in a dilute range, for higher coverage higher order neighbors affect the distribution of atoms, and an indirect correlation term needs to be taken into account. The separation of $g(r)$ into a direct and indirect correlation term is known in the theory of fluids as the Ornstein-Zernike relation and can be solved using the Percus-Yevick approximation \cite{Hansen1976}. A self-consistent determination returns $v(r)$ between pairs of atoms as the quantity of interest \cite{Supplement}.

\begin{figure}[htbp]
	\includegraphics[width=0.6\columnwidth]{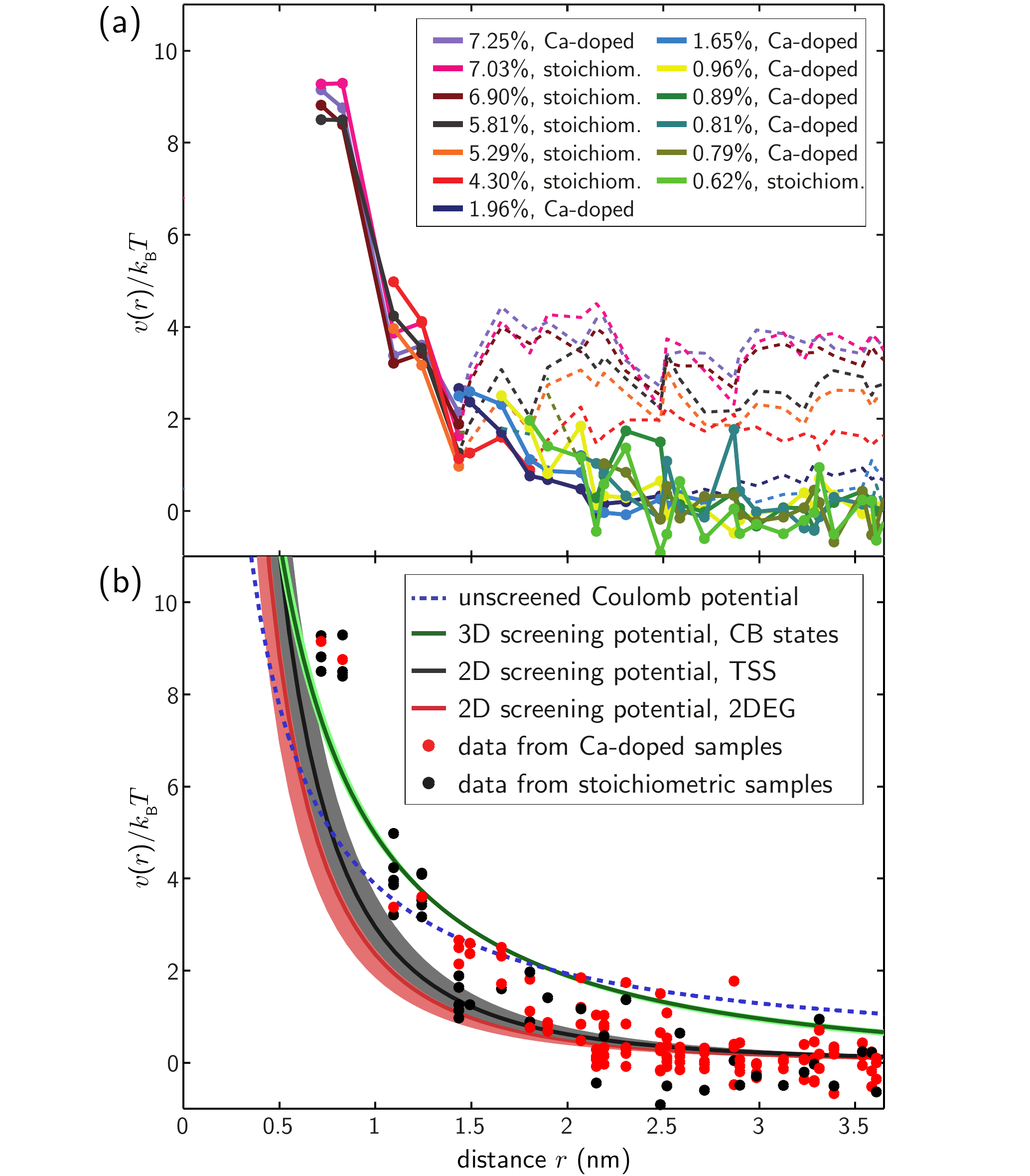}
	\caption{(a) Reduced pair potential, $v(r)/k_{\rm{B}}T$, for various Rb coverages and for stoichiometric and Ca-doped Bi$_2$Se$_3$. The lines connect points at a given coverage as a guide to the eye. The dashed lines mark regions where $v(r)/k_{\rm{B}}T$ is affected by a limitation of the Percus-Yevick approximation.	(b) The experimental data from (a), taking only the data points into account which are connected by solid lines, is fitted with different unscreened and screened Coulomb potentials, where the shaded regions indicate the error bars (see text).}
	\label{E_r_curves}
\end{figure}
In Fig.~\ref{E_r_curves}(a) the reduced pair potential $v(r)/k_\mathrm B T$ is plotted for different surface coverages and different bulk dopings. As indicated in Fig.~\ref{Topographies}, short separations are forbidden and $v(r)$ curves start at $r$ considerably larger than $0$.
For small distances, $v(r)$ decreases monotonically regardless of coverage. However, for larger $r$ there is an increase in $v(r)$ for higher coverage due to the influence of higher order neighbors in $w(r)$. This is an artifact in $v(r)$ due to the breakdown of the Percus-Yevick approximation of $\Delta w(r)$ at larger distances~\cite{Supplement}. 
Hence, the tails of the $v(r)$ curves for the higher coverages that do not asymptotically approach zero [dashed lines in Fig.~\ref{E_r_curves}(a)] are not considered for the further analysis. 

To extract the potential landscape responsible for the repulsive interaction between neighboring charged atoms, we fitted $v(r)/k_{\rm{B}}T$ to various types of Coulomb potentials [Fig.~\ref{E_r_curves}(b)].  Immediately, an unscreened Coulomb potential, where the dielectric constant $\varepsilon_r$ is the sole free parameter, can be excluded. The experimental data decays much faster than the fitted function. 
Therefore, we consider a screened Coulomb potential where free electrons can screen the charged atoms. The screening electrons can originate from three sources: (i) bulk states, (ii) band-bending induced QWS in the conduction band \cite{Bianchi2012}, and (iii) TSS.  In the following, we distinguish between the screening effects resulting from these three different sources within the Thomas-Fermi theory of screening.

The bulk-related screening can be readily ruled out. Because the positively charged Rb atoms can be screened only by negative charges, one would expect a significantly larger screening length for Ca-doped ($p$-doped) Bi$_2$Se$_3$ than for stoichiometric (degenerately $n$-doped) Bi$_2$Se$_3$ if there would be a decisive bulk-related screening. This effect is not observed experimentally, i.e. the data shows no dependence on the bulk doping [Fig.~\ref{E_r_curves}(a)]. Nevertheless, we try to fit the data to a bulk-screening model
	\[v_\mathrm{3D}(r) = \frac{e^2}{4\pi\varepsilon_\mathrm o	(\varepsilon_{r}+1)/2}\frac{1}{r}\,\mathrm e^{-r/R_\mathrm{3D}}\ ,\]
with the screening length \cite{Krcmar20002002, Laubsch2009}
	\[R_\mathrm{3D} = \sqrt{\frac{\varepsilon_\mathrm o \varepsilon_{r}}{e^2\,\partial n/\partial \mu}}\ .\]		
We consider that each Rb atom has a charge of one full electron~\cite{Valla2012, Bianchi2012}.
$\partial n/\partial\mu$ is related to the density of states at $E_{\rm{F}}$, assuming a typical parabolic dispersion for a 3D system \cite{Laubsch2009}.
The value of $E_\mathrm F$ was taken from ARPES measurements of stoichiometric samples \cite{Bianchi2012a}, and the effective mass from \cite{Koehler1973} was used here.
$\varepsilon_r$ is used as the only fitting parameter, yielding~\mbox{$\varepsilon_{r}^{3\mathrm D}=(16.2\pm 1.3)$} and \mbox{$R_\mathrm{3D} = (3.70 \pm 0.61)$~nm}. This value of ~\mbox{$\varepsilon_{r}^{3\mathrm D}$} is unreasonably smaller than the well-characterized and well-accepted bulk value of 113 \cite{Richter1977}. Moreover, the fitted 3D potential in Fig.~\ref{E_r_curves}(b) (green curve) decays slower than the experimental data and poorly fits the data. Clearly, all these observations lead to the conclusion that the bulk-related screening is not decisive in the screening of Rb atoms at the surface, rendering the surface free electrons responsible for the screening potential.

To quantify the 2D potential produced by surface-related free electrons originating from those occupying the QWS and the TSS, we utilize the following form of the potential~\cite{Krcmar20002002, Laubsch2009}, which represents the situation of a 2D conductive sheet on a bulk dielectric with a dielectric constant of $\varepsilon_r$,
	\[v_\mathrm{2D}(r) = \frac{e^2}{4\pi \varepsilon_0 (\varepsilon_{r}+1)/2} \frac{1}{r} \left[1-\frac{\pi}{2} \xi \left(H_0 \left(\xi\right)-N_0 \left(\xi\right)\right)\right]\]
with $\xi = r/R_\mathrm{2D}$, $H_0$ and $N_0$ the Struve and Neumann functions, respectively, and the screening length
	\[R_\mathrm{2D} = \frac{\varepsilon_0(\varepsilon_{r}+1)}{e^2\,\partial n/\partial \mu}\ .\]
To exemplify the role of both electron types on the screening, we only consider the limiting cases where the screening is done solely by either the QWS or the TSS electrons.
$\partial n/\partial \mu$ is calculated assuming a parabolic 2D dispersion \cite{Laubsch2009} in the former, and a linear dispersion in the latter case \cite{Culcer2010}. In both cases, $\partial n/\partial \mu$ depends only weakly on the Rb coverage, as the surface band bending is always close to saturation \cite{Valla2012}. A remaining variation is considered in error bars of the parameters, which are obtained from ARPES measurements \cite{Bianchi2012}.

In figure \ref{E_r_curves}(b), both 2D potentials have been fitted to the experimental data, again using $\varepsilon_{r}$ as only fitting parameter (red and black curve), yielding \mbox{$\varepsilon_{r}^{\rm{QWS}}=(6.8\pm 2.4)$} and \mbox{$R_\mathrm{2D}^{\rm{QWS}} = (0.66 \pm 0.01)$~nm} for the QWS, and \mbox{$\varepsilon_{r}^{\rm{TSS}} = (5.2 \pm 1.2)$} and \mbox{$R_\mathrm{2D}^{\rm{TSS}} = (0.66 \pm 0.01)$~nm} for the TSS, respectively.
The two curves are in better agreement with the experimental data. In particular, the decay of each fit better recounts the experimental behavior.

The fits yield an unexpectedly short $R_\mathrm{2D}$ resulting from the very small \mbox{$\varepsilon_{r}$} as compared to literature values ranging from 30 to 113 \cite{Beidenkopf2011, Richter1977}. The small value of \mbox{$\varepsilon_{r}$} obtained in the fit implies that the screening process is dominated by surface electrons which are just weakly coupled to the underlying substrate bulk. Indeed, the model underlying $v_\mathrm{2D}(r)$ would be a mere two-dimensional metallic sheet in the case of $\varepsilon_r = 1$ \cite{Krcmar20002002}. 
In each fit, we considered the effect of each screening process individually.  However, in reality all screening potentials contribute to the observed Rb distribution, and this may account for deviations in the fits when compared to the experimental data.

In addition to self-assembly of Rb atoms driven by the underlying screening potential, we demonstrate it is possible to atomically manipulate individual Rb atoms on the surface of the topological insulator. Fig.~\ref{Outlook}(a-b) illustrates first the self-assembly of Rb atoms into an ordered array and subsequent manipulation of Rb atoms into the shape of a corral with the STM tip~\cite{Crommie1993}. By combining the ability to do atomic engineering with charged impurities, with the knowledge of the 2D potential landscape, it is possible to tailor a particular energy landscape at the topological surface while simultaneously predicting the resultant 2D electrostatic potential [Fig.~\ref{Outlook}(c)]. Using this technique, patches of a locally high homogeneous Rb-coverage on a Rb-free area of the surface can be assembled. For a Ca-doped substrate, where the TSS is unoccupied without Rb-coverage but occupied below the Rb-patch, this results in a quantum confinement of the Dirac electrons  \cite{Hamalainen2011} of the TSS, which can be tailored by the size and shape of the patch in a very controlled fashion.

In summary, we have mapped the screened Coulomb potential on a topological surface driven by charged Rb atoms. Our observations reveal that screening of surface electrons at Bi$_2$Se$_3$(111) is surprisingly efficient as manifested by a short screening length. 
While the charge screening here is an interplay between different screening potentials, the role of the TSS on charge screening could be further understood by comparing these results with complementary studies of alkali atom distributions on other prototypical TI systems like Bi$_2$Se$_2$Te or Bi$_2$Te$_3$ where the Fermi surface is different compared to Bi$_2$Se$_3$.
Combining the electrostatic screening around the charged impurities and the ability to perform atomic-scale manipulation, we have demonstrated the ability to engineer the potential landscape at the surface of a topological insulator.

We acknowledge financial support from the DFG via SFB 668, Graduiertenkolleg 1286, and SPP 1666, from the ERC Advanced Grant ``FURORE'', from the VILLUM foundation, and from the Danish National Research Foundation. A.A.K. acknowledges Project No. KH324/1-1 from the Emmy-Noether-Program of the DFG.

\end{document}